\documentclass[final,1p,times]{elsarticle} % do not change
\usepackage{graphicx} % do not change
\usepackage{amssymb} % do not change
\usepackage{amsthm} % do not change
\usepackage{lineno} % do not change

\journal{Nuclear Physics A} % do not change
\begin{document} % do not change

\begin{frontmatter} % do not change

%% QM09Author: please enter your  
%% Title, author and address info here; please do not use footnotes

% Your Title - please modify
\title{Two Ridges, One Explanation}

% Principle author, and co-authors - please modify
\author{George Moschelli$^{a}$ and Sean Gavin$^{a}$}

% Address - please modify
% note that if you have authors from several institutions, we recommend
% labelling these [a], [b], [c] etc.
\address[a]{Physics Department, Wayne State University, % label [a]
%666 W. Hancock,
Detroit, MI, 48202, USA}

\begin{abstract} % do not change
We attribute the phenomenon known as ``the ridge"  to long range initial state correlations from Color Glass Condensate flux tubes and later stage radial flow. We show that this description can explain the amplitude and azimuthal width of the soft ridge and nearly explain that of the hard ridge, suggesting that the two are essentially the same phenomenon.
\end{abstract} % do not change

\end{frontmatter} % do not change

%% QM09: we keep linenumbers at least for initial version
%\linenumbers % do not change

%% start of main text - please modify. Below is a sub-set (single section) 
%% of an earlier proceedings that show how one can handle references 
%% and figures etc.
%%\section{}\label{}

%\section{Correlations and Transverse Momentum Dependence}
%
The phenomenon known as ``the ridge" has been measured both with and without a jet trigger \cite{Putschke:2007mi,Daugherity:2006hz}. In \cite{Long} we propose an explanation for the untriggered ``soft ridge" that depends on transverse flow building on early stage spatial correlations due to CGC flux tubes. Here, we study the $p_t$ dependence of the ridge by examining the behavior of our soft ridge calculation when the $p_t$ range is increased. The ``hard ridge" has been thought of as a separate phenomenon since the $p_t$ of correlated pairs is chosen in a narrow range with the intent to examine the effects of jets on the medium.  We show that the contribution of correlations of soft particles in this range is the dominant factor and that the hard and soft ridges are essentially the same phenomenon.

In \cite{Long} we quantitatively describe the amplitude and azimuthal width for both 200 and 62 GeV measurements in Au+Au. The key features of the model are that CGC theory allows for early stage correlations to extend over several units of rapidity while predicting the centrality and energy dependence of the gluon multiplicity  \cite{larry,Kharzeev:2000ph,mv,lappi}, and that transverse flow enhances angular correlations \cite{Voloshin:2003ud}.
We study the $p_t$ dependence of the soft ridge in the context of \cite{Long} by increasing the lower $p_t$ limit of correlated particles toward the hard ridge range. We find that soft, or bulk-bulk, correlations in the $p_t$ range of the hard ridge nearly explain the data. To completely describe the data, we include correlations of jets with bulk particles following \cite{Shuryak:2007fu}.

\begin{figure}[ht]
\centering
\includegraphics[width=0.4\textwidth]{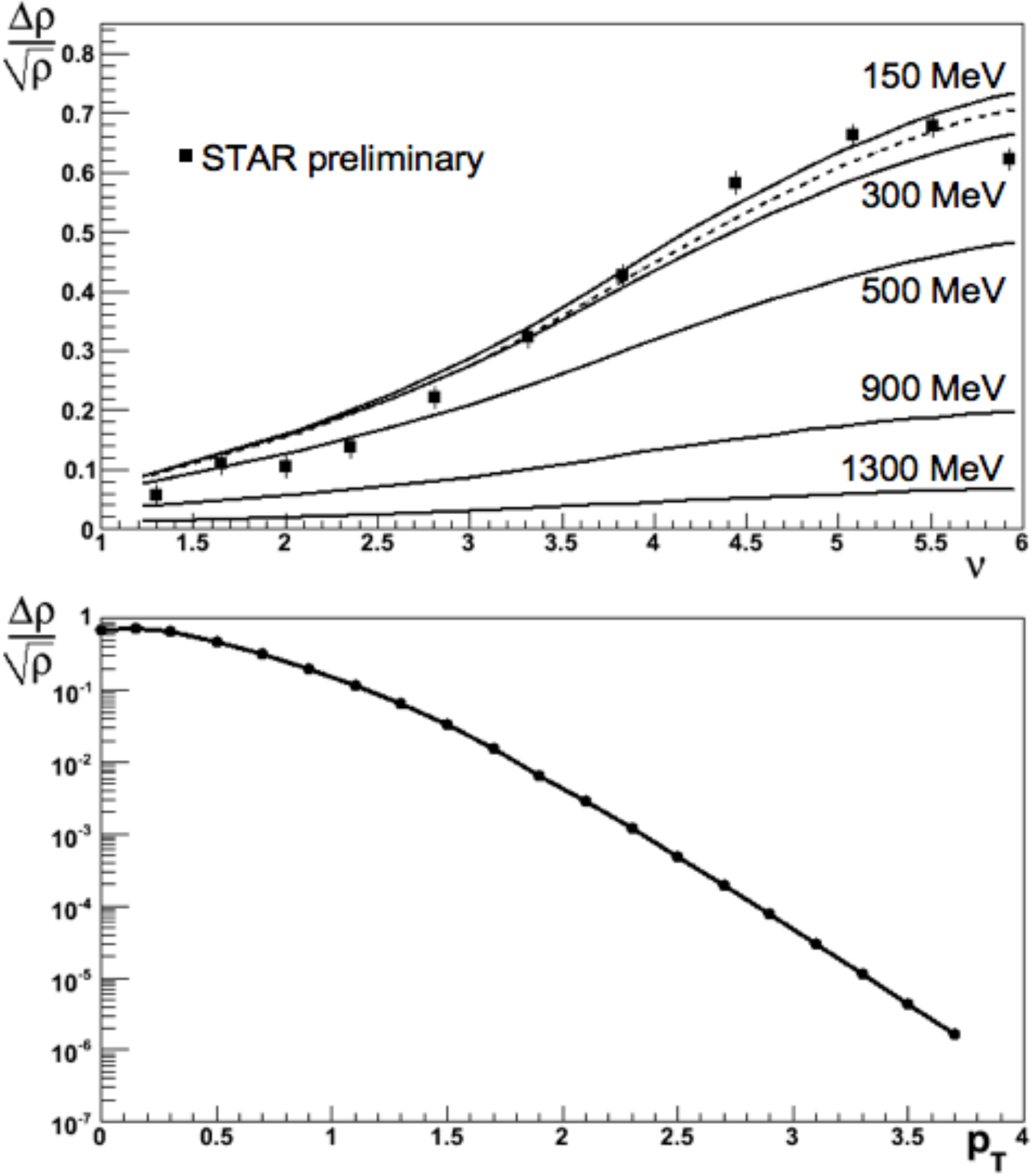}
\includegraphics[width=0.45\textwidth]{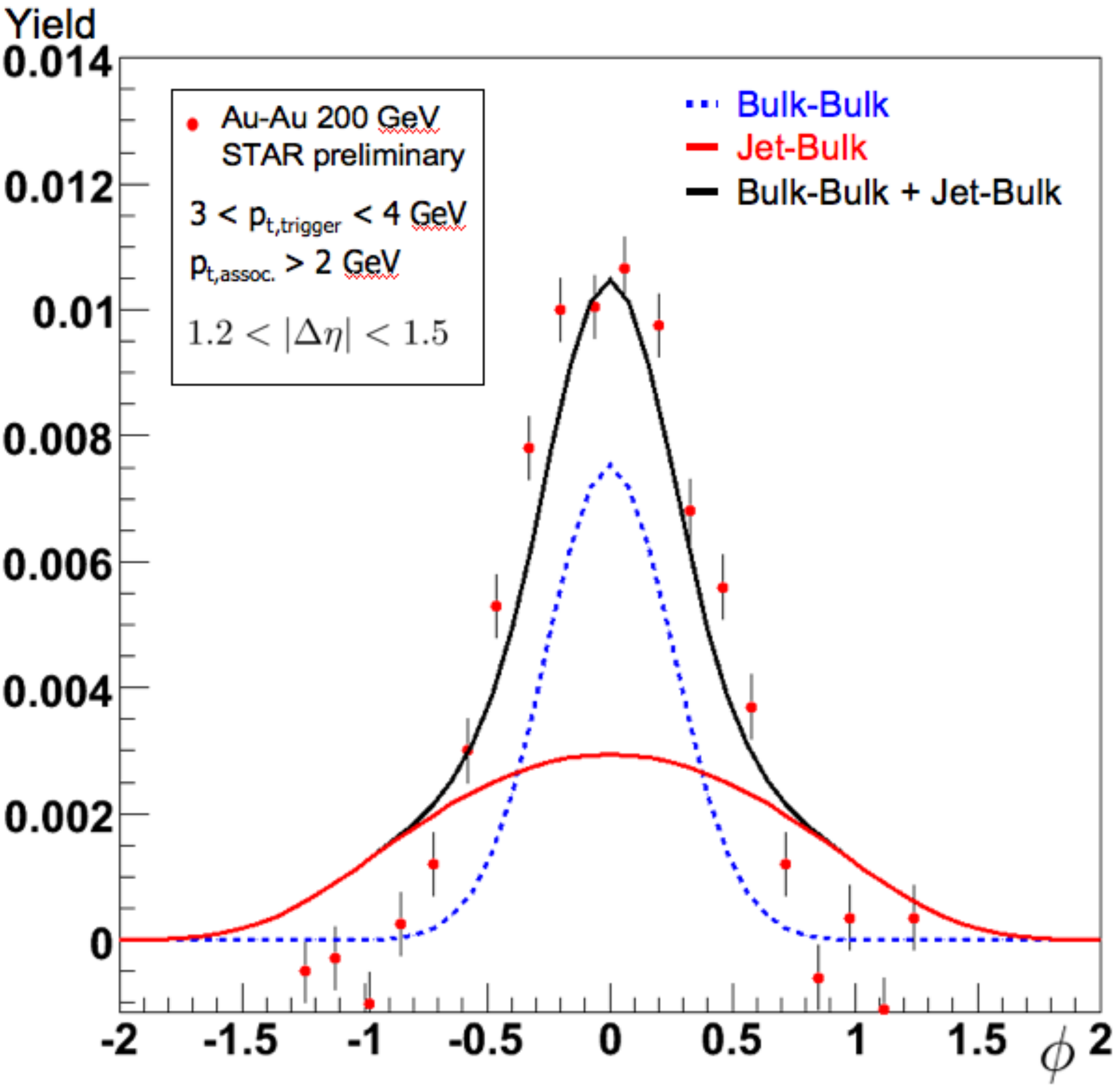}
\caption[]{For Au+Au 200 GeV. Upper left: bulk-bulk correlations vs. centrallity for different $p_{t~min}$ ($p_{t~min}=0$ dashed line). Lower Left: most central point of the bulk-bulk curves vs.$p_{t~min}$ of that curve. Right: bulk-bulk and jet-bulk contributions to the hard ridge}
\label{fig1}
\end{figure}

In \cite{Long} we study the quantity $\Delta\rho/\sqrt{\rho_{ref}}$ integrated over $p_t$. This would correspond to the dashed curve in the upper left panel of Fig.\ref{fig1}.
We extend this here by considering $\left[\Delta\rho/\sqrt{\rho_{ref}}\right]_{p_{t~min}}$ integrated over the range $p_{t~min}<p_{t,1},p_{t,2}<\infty$. The remaining curves in the upper left panel correspond to different $p_{t~min}$ limits. The $p_{t~min}$ limit is increased to the range where jets are expected to dominate the spectrum. Although the magnitude of the correlations decrease, they are still non-zero. The behavior of this decrease is illustrated in the lower left panel of Fig.\ref{fig1}, where the most central point of $\left[\Delta\rho/\sqrt{\rho_{ref}}\right]_{p_{t~min}}$ is plotted vs. $p_{t~min}$. To directly compare to the hard ridge, we integrate our correlation function with $p_t$ limits corresponding to the trigger and associated ranges and  convert $\Delta\rho/\sqrt{\rho_{ref}}$ to $yield$. The result is the dashed blue curve in the right panel of Fig.\ref{fig1}. One can see that bulk-bulk correlations alone are a significant fraction of the data and slightly narrower.

We also calculate the contribution to the hard ridge from jet correlations with flux tubes. We follow \cite{Shuryak:2007fu} with the caveat that we correlate the radial position of the hard collision with a flux tube at the same position. The result is the red curve in the right panel of Fig.\ref{fig1}. Again, one can see that the jet-bulk contribution contributes less to the amplitude and is wider than the data. The addition of the jet-bulk and bulk-bulk yields however give nice agreement with both the amplitude and azimuthal width of the data as shown by the black curve. 

Although jets dominate the spectrum at higher $p_t$, jet-jet correlations (correlations of jet particles with other jet particles or fragments) could only exist within $\sim 1$ unit of rapidity of each other. The data shown in the right panel of Fig.\ref{fig1} is for correlations $\eta>1$ unit of rapidity apart. It is natural then, that correlations with flux tubes can also explain the long range features of the hard ridge as well as the soft ridge.

%% end of main text

\section*{Acknowledgments} % please check/modify, comment out or delete if not needed
We thank R. Bellwied, C. DeSilva, and J. Nagle.
This work was supported in part by U.S. NSF PECASE/CAREER grant PHY-0348559.

 % do not change 
\end{document}